\documentclass[9pt,twocolumn,twoside]{osajnl}

\journal{ol} 

\setboolean{shortarticle}{true}

\usepackage{graphicx}

\usepackage{lineno}
\usepackage{graphicx}
\usepackage[caption=false, font=normalsize, labelfont=sf, textfont=sf]{subfig}
\usepackage[justification=centering]{caption}

\title{On-chip low-loss all-optical MoSe$_{2}$ modulator}

\author[1,*]{Mohammed Alaloul}
\author[2]{Jacob B. Khurgin}
\author[1]{Ibrahim Al-Ani}
\author[1]{Khalil As'ham}
\author[3]{Lujun Huang}
\author[1]{Haroldo T. Hattori}
\author[1,*]{Andrey E. Miroshnichenko}

\affil[1]{School of Engineering and Information Technology, University of New South Wales, Canberra, ACT 2612, Australia}
\affil[2]{Electrical and Computer Engineering Department, Johns Hopkins University, Baltimore, MD 21218, USA}
\affil[3]{School of Physics and Electronic Science, East China Normal University, Shanghai 200241, China}

\affil[*]{Corresponding authors: m.alaloul@unsw.edu.au; andrey.miroshnichenko@unsw.edu.au}

\begin{abstract}
Monolayer transition metal dichalcogenides (TMDCs), like MoS$_2$, MoSe$_2$, WS$_2$, and WSe$_2$, feature direct bandgaps, strong spin-orbit coupling, and exciton-polariton interactions at the atomic scale, which could be harnessed for efficient light emission, valleytronics, and polaritonic lasing, respectively. Nevertheless, to build next-generation photonic devices that make use of these features, it is first essential to model the all-optical control mechanisms in TMDCs. Herein, a simple model is proposed to quantify the performance of a 35$\,$\textmu m long Si$_3$N$_4$ waveguide-integrated all-optical MoSe$_2$ modulator. Using this model, a switching energy of 14.6$\,$pJ is obtained for a transverse-magnetic (TM) and transverse-electric (TE) polarised pump signals at $\lambda =\,$480$\,$nm. Moreover, maximal extinction ratios of 20.6$\,$dB and 20.1$\,$dB are achieved for a TM and TE polarised probe signal at $\lambda =\,$500$\,$nm, respectively, with an ultra-low insertion loss of $<0.3\,$dB. Moreover, the device operates with an ultrafast recovery time of 50$\,$ps, while maintaining a high extinction ratio for practical applications. These findings facilitate modeling and designing novel TMDC-based photonic devices.
\end{abstract}

\setboolean{displaycopyright}{true}

\begin{document}

\maketitle

Two-dimensional (2D) materials have recently emerged as practical active materials for on-chip photonics \cite{sun2016optical, akinwande2019graphene, alaloul2021plasmonic}. Unlike bulk materials, these materials are easily integrated into underlying substrates by van der Waals (vdW) forces, which facilitate device fabrication by complementary metal-oxide semiconductor (CMOS) processes \cite{alaloul2021plasmon}. Moreover, these materials exhibit unique optoelectronic properties that are of interest from a scientific and technological perspective. In particular, monolayer transition metal dichalcogenides (TMDCs) have direct bandgaps, strong spin-orbit coupling, and exciton-polariton interactions at the atomic scale, which could be utilized for efficient light emission, valleytronics, and polaritonic lasing, respectively \cite{huang2021enhanced, rauschenbeutel2022chiral, kavokin2022polariton, ou2021spatial}. To design and build functional devices that make use of these features, it is first essential to model the electrical and all-optical control mechanisms in TMDCs. So far, there has not been a comprehensive model that describes these mechanisms in TMDC monolayers. To fill this gap, we model the saturable absorption and all-optical modulation mechanisms in monolayer molybdenum diselenide (MoSe$_2$), which is integrated into silicon nitride (Si$_3$N$_4$) waveguides. To the best of our knowledge, a waveguide-integrated all-optical MoSe$_2$ modulator has not been reported in the literature. This model can also describe these mechanisms in other TMDCs and 2D semiconductors. 

\begin{figure} 
    \centering
  \subfloat[\label{1a}]{%
       \includegraphics[width=0.65\linewidth]{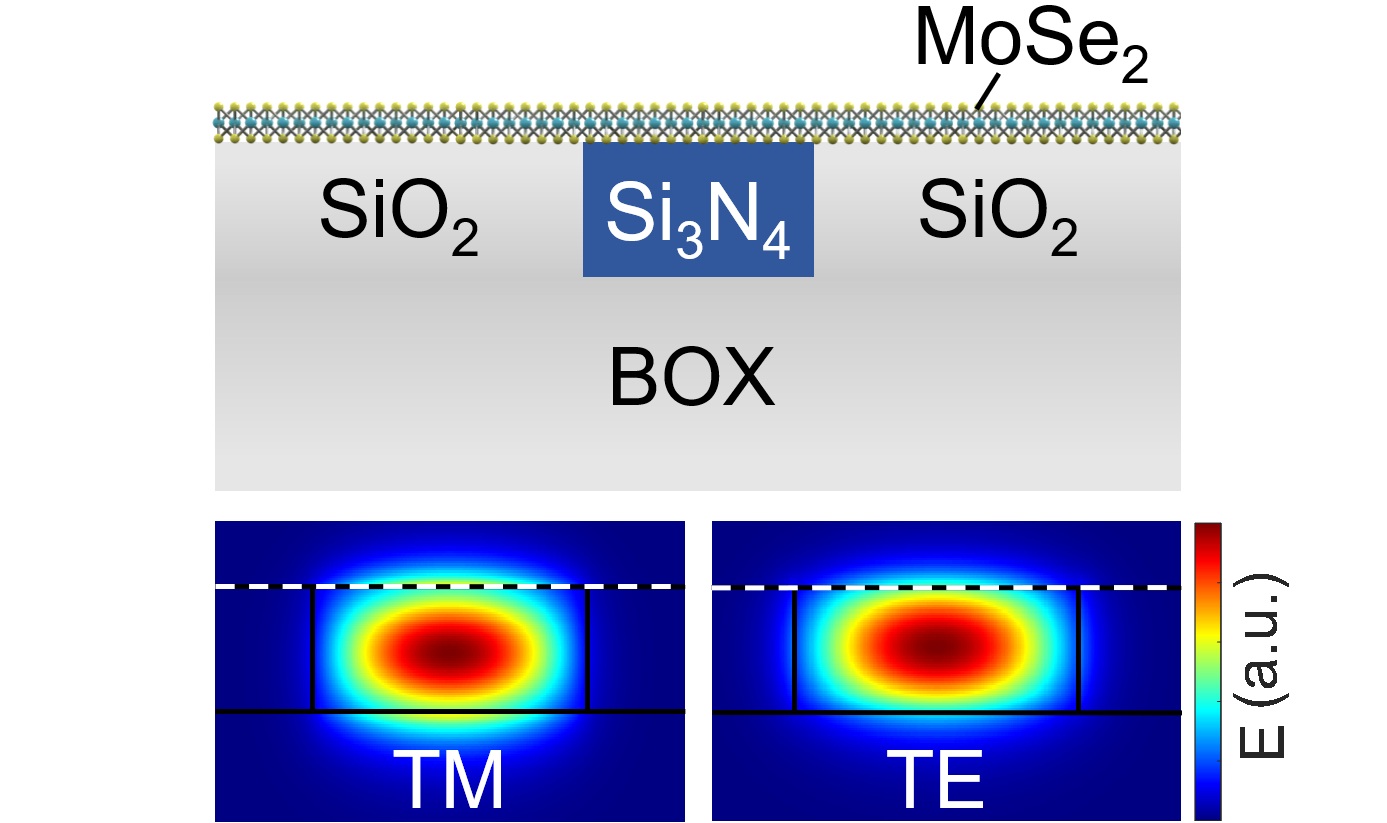}}
    \hfill
  \subfloat[\label{1b}]{%
        \includegraphics[width=0.9\linewidth]{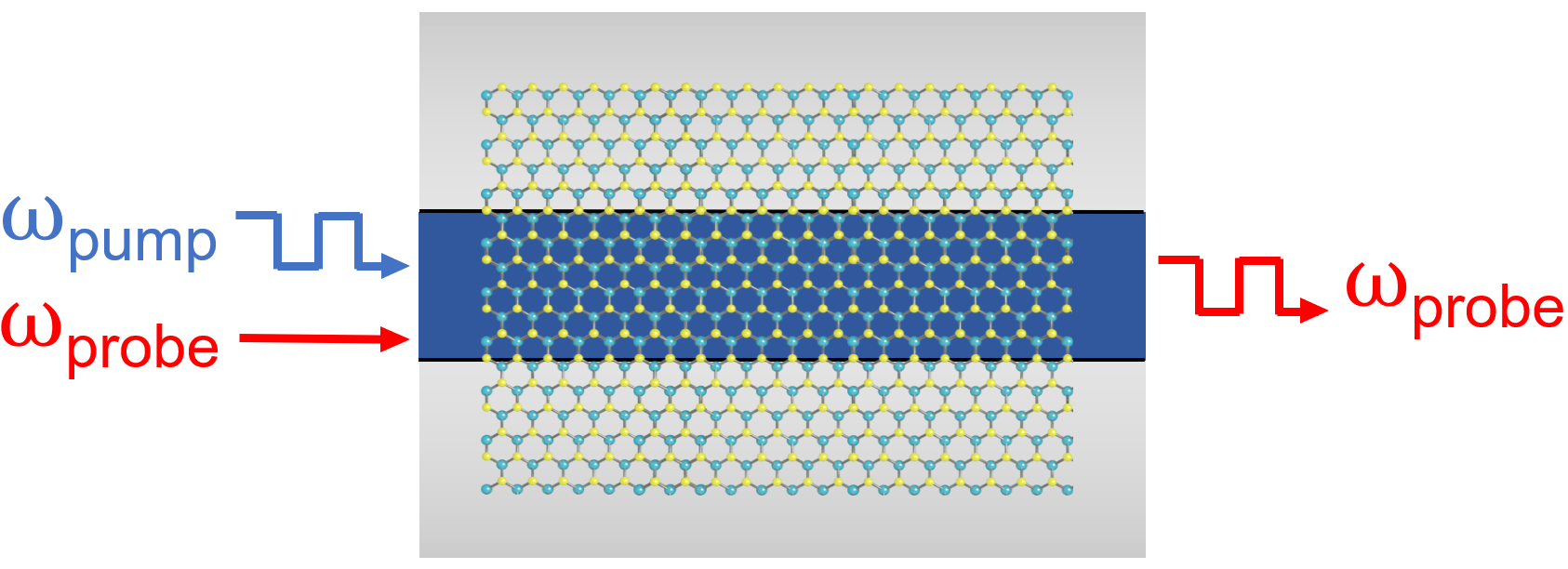}}
  \caption{(a) Cross-sectional view of the on-chip modulator. Bottom panel: electric field profile ($E$) of the TM and TE modes at $\lambda = 450\,$nm. The white dashed line represents the MoSe$_2$ plane. (b) Top view. The pump signal modulates a probe signal. MoSe$_2$: molybdenum diselenide, Si$_3$N$_4$: silicon nitride, SiO$_2$: silicon dioxide, BOX: buried oxide.}
  \label{fig1} 
\end{figure}

\begin{figure} 
    \centering
  \subfloat[\label{length_absorption}]{%
       \includegraphics[width=0.5\linewidth]{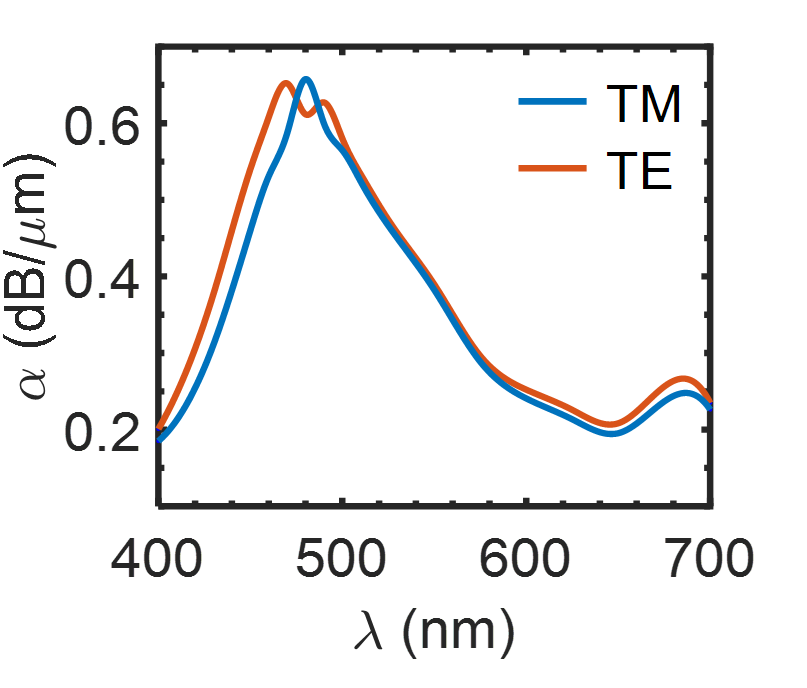}}
    \hfill
  \subfloat[\label{layers_absorption}]{%
        \includegraphics[width=0.5\linewidth]{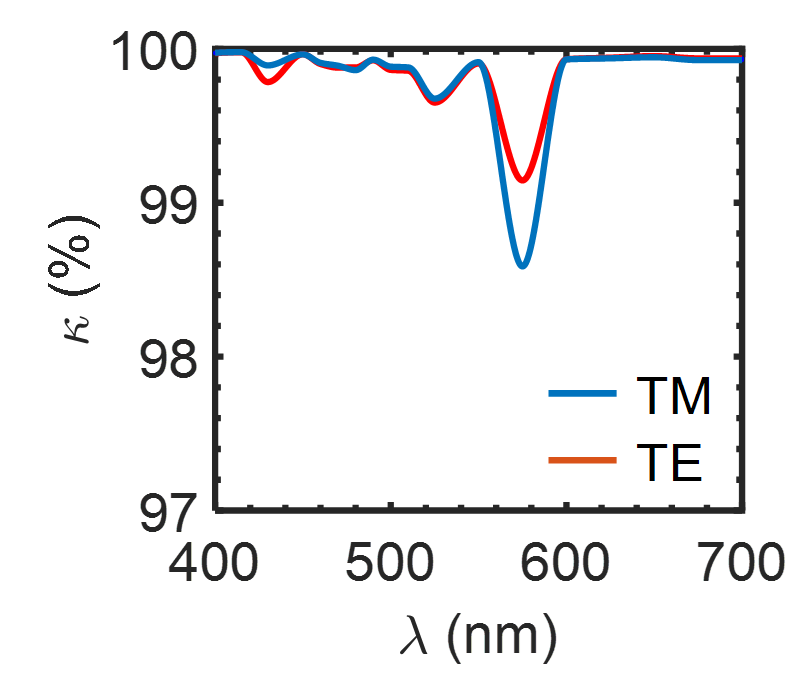}}
  \caption{(a) Absorption coefficients ($\alpha$) and coupling efficiencies ($\kappa$) of the TM and TE modes as a function of wavelength ($\lambda$).}
  \label{fig2} 
\end{figure}

\begin{figure*}[t]
    \centering
    \includegraphics[width=0.9\textwidth]{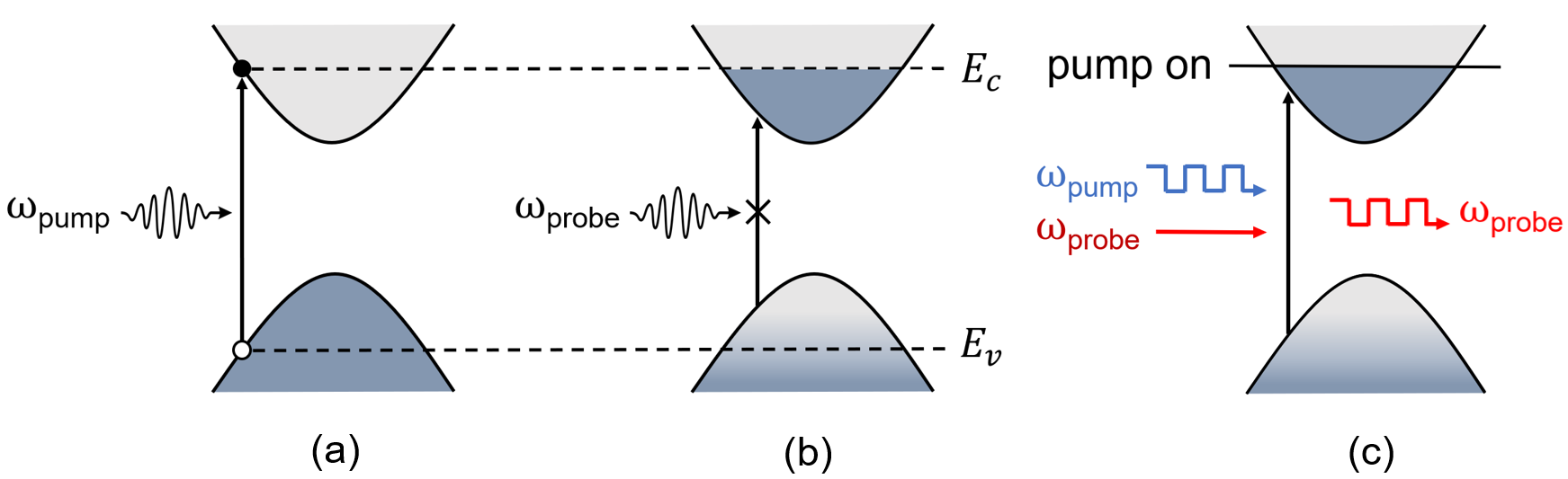}
    \caption{(a) Interband absorption of a pump photon with an energy $\hbar \omega_{\text{pump}}$. (b) Probe photon transmitted after applying a pump signal with sufficiently high intensity. (c) Transmission of the probe signal, as determined by the pump signal amplitude. Black and white circles represent electrons and holes, respectively. Darker shades represent filled energy states.}
    \label{fig:transitions}
\end{figure*}

The device structure is presented in Fig. \ref{fig1}. A $750\times 250\,$nm Si$_3$N$_4$ wire waveguide on top of a $10\,$\textmu m thick buried oxide (BOX) layer guides the incoming light to the modulator section, where the MoSe$_2$ monolayer is located. The waveguide dimensions are optimized to achieve an ultra-low propagation loss. Si$_3$N$_4$ waveguides with similar dimensions have been demonstrated in \cite{buchberger2021modelling}, where the optical mode is guided with a propagation loss of merely 1.7$\,$dB/cm. The waveguide supports the transverse-magnetic (TM) and transverse-electric (TE) modes. The computed absorption coefficient ($\alpha$) of each mode is presented in Fig. \ref{length_absorption}, which is attributed to the MoSe$_2$ monolayer because it is the only absorbing material within the structure. Highly efficient coupling ($>98$\%) between the Si$_3$N$_4$ waveguide and the modulator section is obtained for both modes, as shown in Fig. \ref{fig2}b. Simulations were conducted in Lumerical MODE with a 0.7$\,$nm thick MoSe$_2$ monolayer. The refractive index data of monolayer MoSe$_2$ are taken from \cite{hsu2019thickness}. Perfect matching layer (PML) boundary conditions were used in the simulations. Similar results are obtained using metal boundary conditions. 

The modulation relies on the exciton bleaching mechanism \cite{amin2018waveguide}. As depicted in Fig. \ref{fig:transitions}a, the MoSe$_2$ monolayer absorbs an incident pump signal photon, which generates an interband exciton at visible wavelengths. When a sufficiently intense pump signal is applied, exciton bleaching occurs because of electron-hole plasma screening or band filling \cite{chemla1985room}. Due to the reduced dimensionality in 2D materials, screening is reduced. Then, exciton bleaching is most likely due to band filling \cite{amin2018waveguide}. Taking that into account, when the generated excitons fill the conduction and valence band states, a probe signal photon with an energy $\hbar \omega_{\text{probe}} \leq \hbar \omega_{\text{pump}}$ is transmitted, as shown in Fig. \ref{fig:transitions}b. By taking advantage of this phenomenon, amplitude modulation is realized \cite{alaloul2022high}, where the probe signal is HIGH when the pump signal is ON, whereas it is LOW when the pump signal is OFF, as shown in Fig. \ref{fig:transitions}c. Herein, the band structure of monolayer MoSe$_2$ is described by the dispersion relations of the conduction ($E_c$) and valence ($E_v$) bands:

\begin{equation}
    E_c (k) = \dfrac{E_g}{2} + \dfrac{\hbar^2 k^2}{2 m_c}
\end{equation}
    
\begin{equation}
    E_v (k) = -\dfrac{E_g}{2} - \dfrac{\hbar^2 k^2}{2 m_v}
\end{equation}

\noindent where $E_g = 1.55\,$eV is the energy bandgap \cite{tongay2012thermally}, $m_c = 0.545$ and $m_v = 0.647$ are the effective masses of electrons and holes in monolayer MoSe$_2$ \cite{rawat2018comprehensive}, respectively, and $k = \sqrt{2 \pi n}$ is the wavevector, with $n$ being the carrier density. For a pump signal photon with an energy $\hbar \omega_{\text{pump}}$, transitions are blocked when the photogenerated excitons fully occupy the states. From Fig. \ref{fig:transitions}, this occurs when:

\begin{equation}
   \hbar \omega_{\text{pump}} = E_c (k) - E_v (k) = E_g + \dfrac{\hbar^2 k^2}{2 m_r}
\end{equation}

\noindent where $m_r = (1/m_{c} + 1/m_{v})^{-1}$ is the reduced effective mass. The corresponding photogenerated exciton density is given by:

\begin{equation}
    n = \dfrac{m_r}{\pi \hbar^2} (\hbar \omega_{\text{pump}} - E_g)
\end{equation}

Then, the switching energy of the modulator ($U_{\text{sw}}$) is calculated by multiplying the  total number of photogenerated excitons ($m$) by $\hbar \omega_{\text{pump}}$ \cite{alaloul2021low}:

\begin{equation} \label{switchingEqation}
U_{\text{sw}} = \sum_{m} \hbar\omega_{\text{pump}}
\end{equation}

\noindent where $m = n W L_{\text{mod}}$ is the total number of photogenerated excitons, $W$ is the waveguide width, and $L_{\text{mod}}$ is the modulator length. For $L_{\text{mod}} = 35\,$\textmu m, $\sim99\,$\% of the pump signal is absorbed for both the TM and TE polarisations, where the absorbed power fraction ($A$) is calculated using the Beer-Lambert Law $A (L_{\text{mod}}) = 1 - 10^{-(\alpha /10)*L_{\text{mod}}}$. A fraction of the absorbed light power is non-saturable ($A_{\text{ns}}$), and does not contribute to the modulation mechanism. Besides, there is a small coupling loss, $\Gamma = 1 - \kappa$, between the Si$_3$N$_4$ waveguide and the modulator section. To account for these losses, the effective switching energy ($U_{\text{eff}}$) is calculated as \cite{alaloul2022high}:

\begin{equation} \label{ueff_energy}
    U_{\text{eff}} = \dfrac{U_{\text{sw}}(1 + \Gamma + A_{WG})}{A*(1-A_{\text{ns}})}
\end{equation}

\noindent where $A_{\text{ns}} = (1.08/25.58)*100\% = 4.22\%$ is taken based on the values reported for a MoSe$_2$ monolayer in \cite{tian2022broadband}. $A_{WG} = 1-10^{-(\alpha_{WG}/10)*L_{\text{mod}}}$ is the Si$_3$N$_4$ waveguide loss that is not related to MoSe$_2$, where $\alpha_{WG} = 1.7\,$dB/cm. In Fig. \ref{length_absorption}), the absorption peak of the waveguide-integrated MoSe$_2$ occurs at $\sim 480\,$nm, which is chosen as the pump signal wavelength. The resulting $U_{\text{eff}}$ is $\sim14.6 \,$pJ for both the TE- and TM-polarised modes, where the absorption coefficients of both modes at that wavelength are almost similar. 

\begin{figure} 
    \centering
  \subfloat[\label{ER}]{%
       \includegraphics[width=0.5\linewidth]{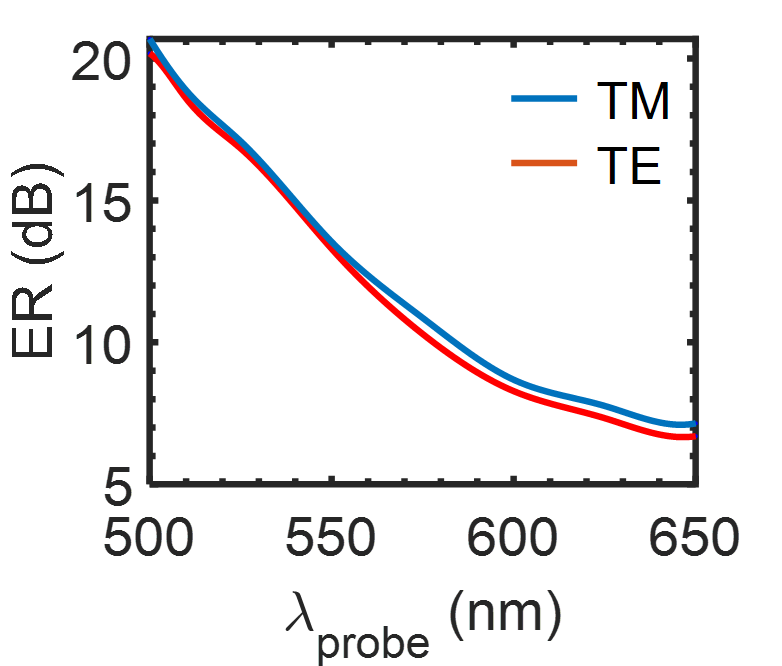}}
    \hfill
  \subfloat[\label{IL}]{%
        \includegraphics[width=0.5\linewidth]{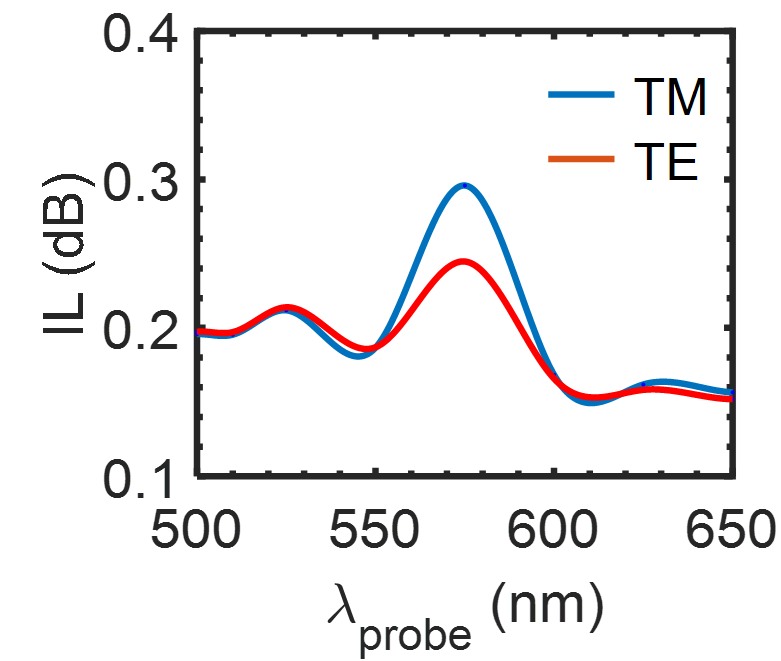}}
  \caption{(a) Extinction ratio ($ER$) and (b) insertion loss ($IL$) as a function of the probe signal wavelength ($\lambda_{\text{probe}}$) for the TM and TE modes.}
  \label{fig4} 
\end{figure}

The modulation efficiency is quantified by the extinction ratio ($ER$), where $ER = 10\text{log}_{10} (T_{\text{on}}/T_{\text{off}})$ \cite{alaloul2022high}. $T_{\text{on}}$ and $T_{\text{off}}$ represent the transmittance of the probe signal when the pump signal is switched ON and OFF, respectively. Here, we consider operation where a pump signal with an energy $U\geq U_{\text{eff}}$ is applied. Then, $T_{\text{on}}$ and $T_{\text{off}}$ are given by \cite{alaloul2022electrical}:

\begin{equation} \label{Tmax}
    T_{\text{on}} = [1 - (\Gamma + A*A_{\text{ns}})]* (1-\Gamma)
\end{equation}

\begin{equation} \label{Toff}
    T_{\text{off}} = [1 - (\Gamma + A)] * (1-\Gamma)
\end{equation}

\noindent where $A$ is the absorbance calculated using the Beer-Lambert law. In addition, the insertion loss ($IL$) of the device is calculated as $IL = 10 \text{log}(1/T_{\text{on}})$. The resulting $ER$ and $IL$ of the device are presented in Fig. \ref{fig4} for a broadband wavelength region. First, it is found that the $IL$ is as low as $<0.3\,$dB across the studied band, which is attributed to the ultra-low coupling loss of the modulator. A remarkable $ER$ ratio of 20.6$\,$dB and 20.1$\,$dB are achievable for a 500$\,$nm probe signal with the TM and TE polarisations, respectively. Moreover, the $ER$ plot follows the trend of the absorption coefficient plot that was presented in Fig. \ref{fig2}a. A higher absorption of the probe signal results in a greater extinction ratio between $T_{\text{on}}$ and $T_{\text{off}}$. Additionally, highly efficient modulation is attainable at other wavelengths, where the lowest $ER$ is 7.1$\,$dB and 6.7$\,$dB for the TM and TE polarised modes, respectively, at $\lambda_{\text{probe}} = 650\,$nm.

\begin{figure} 
    \centering
  \subfloat[\label{fig:cmin}]{%
       \includegraphics[width=0.5\linewidth]{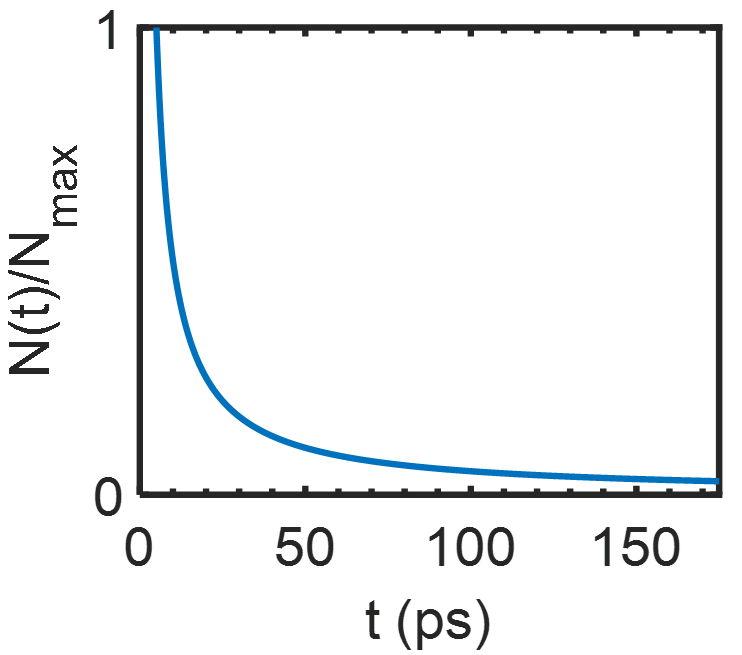}}
  \caption{Normalized exciton density as a function of time.}
  \label{fig:results3} 
\end{figure}

The modulation performance of the device is characterized by its recovery time, which is fundamentally limited by the cooling mechanisms in the MoSe$_2$ monolayer. Following photoexcitation by a pump signal, exciton-exciton annihilation occurs within a few tens of picoseconds, while full electron-hole recombination can last for hundreds of picoseconds \cite{liu2017coupled}. The cooling dynamics in monolayer MoSe$_2$ can be modelled as \cite{liu2017coupled}:

\begin{equation} \label{eq:fit}
    N(t) = \dfrac{N_0}{1 + \gamma N_0 t}
\end{equation}

\noindent where $N(t)$ is the exciton density as a function of time, $N_0$ is the initial exciton density, and $\gamma = 0.4\,$cm$^2$/s is the exciton-exciton annihilation rate \cite{liu2017coupled}. At $\lambda_{\text{pump}} = 480\,$nm, the photogenerated exciton density is $1.28 \times 10^{14}\,$cm$^{-2}$. Using these parameters, Eq. \ref{eq:fit} is plotted in the time range of $5 - 175\,$ps in Fig, \ref{fig:results3}, which is similar to the one presented in \cite{liu2017coupled}. It is observed that recovery mostly occurs within tens of picoseconds by the exciton-exciton annihilation process. As previously explained, photogenerated excitons fill the states, leading to exciton bleaching and transmission of the probe signal. Therefore, the transmission of the probe signal follows the trend of the plotted curve in Fig. \ref{fig:results3}. Even though full recombination occurs in hundreds of picoseconds, for practical switching applications, it would be sufficient to operate the modulator with a time interval of 50$\,$ps following a pump excitation, where the corresponding $N(t)/N_{\text{max}} = 10\,$\%. In compensation for faster operation, the $ER$ decreases because of the reduced contrast between the ON and OFF states. For instance, for a negligible coupling and non-saturable losses, $ER = 20 \,$dB, where $T_{\text{on}} = 100$\%, and $T_{\text{off}} = 1$\%. However, while operating at a 50$\,$ps time interval, $T_{\text{on}} = 100$\%, and $T_{\text{off}} = 10$\%, resulting in a 10$\,$dB extinction ratio, which is highly desirable for practical applications. The ultrafast response of this device could be harnessed for building saturable absorbers, pulsed lasers, and for visible light switches. Moreover, the proposed model can describe the saturable absorption and all-optical modulation mechanisms in other TMDCs and 2D semiconductors by tuning the modelling parameters, e.g., material absorption, bandgap, effective masses, and exciton-exciton annihilation rate.

\begin{backmatter}
\bmsection{Funding} Australian Research Council (DP200101353).

\bmsection{Acknowledgments} This work was supported by the Australian Research Council (DP200101353).

\bmsection{Disclosures} The authors declare no conflicts of interest.

\bmsection{Data Availability Statement} The authors confirm that the data supporting the findings of this study are available within the article.

\end{backmatter}

\bibliography{sample}

\bibliographyfullrefs{sample}


\ifthenelse{\equal{\journalref}{aop}}{%
\section*{Author Biographies}
\begingroup
\setlength\intextsep{0pt}
\begin{minipage}[t][6.3cm][t]{1.0\textwidth} 
  \begin{wrapfigure}{L}{0.25\textwidth}
    \includegraphics[width=0.25\textwidth]{john_smith.eps}
  \end{wrapfigure}
  \noindent
  {\bfseries John Smith} received his BSc (Mathematics) in 2000 from The University of Maryland. His research interests include lasers and optics.
\end{minipage}
\begin{minipage}{1.0\textwidth}
  \begin{wrapfigure}{L}{0.25\textwidth}
    \includegraphics[width=0.25\textwidth]{alice_smith.eps}
  \end{wrapfigure}
  \noindent
  {\bfseries Alice Smith} also received her BSc (Mathematics) in 2000 from The University of Maryland. Her research interests also include lasers and optics.
\end{minipage}
\endgroup
}{}

\end{document}